\documentclass[11pt]{article}
\usepackage{amssymb}
\usepackage{axodraw}

\newcommand{\bi}{\bibitem}
\newcommand{\be}{\begin{eqnarray}}
\newcommand{\ee}{\end{eqnarray}}
\newcommand{\rar}{\rightarrow}

\begin{document}


\begin{titlepage}

\title{A Black Hole Conjecture and Rare Decays in Theories with 
Low Scale Gravity}

\author{C. Bambi$^{a,b,}$\footnote{E-mail: bambi@fe.infn.it},
A.D. Dolgov$^{a,b,c,}$\footnote{E-mail: dolgov@fe.infn.it}
and K. Freese$^{d,}$\footnote{E-mail: ktfreese@umich.edu}}

\maketitle

\begin{center}
$^{a}$Istituto Nazionale di Fisica Nucleare, Sezione di Ferrara,
       I-44100 Ferrara, Italy\\
$^{b}$Dipartimento di Fisica, Universit\`a degli Studi di Ferrara,
       I-44100 Ferrara, Italy\\
$^{c}$Institute of Theoretical and Experimental Physics,
       113259, Moscow, Russia\\
$^{d}$Michigan Center for Theoretical Physics, Physics Dept., University
       of Michigan,\\ Ann Arbor, MI 48109
\end{center}

\vspace{0.5cm}

\begin{abstract}
In models with large extra dimensions, where the fundamental gravity
scale can be in the electroweak range, gravitational effects in
particle physics may be noticeable even at relatively low energies.
In this paper, we perform simple estimates of the decays of
elementary particles with a black hole intermediate state.  Since
black holes are believed to violate global symmetries, particle decays
can violate lepton and baryon numbers. Whereas previous literature has
claimed incompatibility between these rates (e.g. $p$-decay) and existing
experimental bounds, we find suppressed baryon and lepton-violating
rates due to a new conjecture about the nature of the virtual black
holes.  We assume here that black holes lighter than the (effective) 
Planck mass must have zero electric and color charge and zero angular 
momentum -- this statement is true in classical general relativity and 
we make the conjecture that it holds in quantum gravity as well. If true, 
the rates for proton-decay, neutron-antineutron oscillations, and 
lepton-violating rare decays are suppressed to below experimental bounds 
even for large extra dimensions with TeV-scale gravity. Neutron-antineutron 
oscillations and anomalous decays of muons, $\tau$-leptons, and $K$ and 
$B$-mesons open a promising possibility 
to observe TeV gravity effects with a minor increase of existing 
experimental accuracy. 
\end{abstract}

\end{titlepage}


\section{Introduction}

In physics, we sometime run into the so-called hierarchy
problems. This happens when we find two or several quantities which
are different by many orders of magnitude whereas, {\it a priori}, we
would have expected them to be more or less at the same scale.

A very well known hierarchy problem in high energy physics is related
to the huge gap between the Planck mass $M_{Pl} \sim 10^{19}$ GeV,
which sets the energy scale for gravitational interactions, and the
electroweak scale $M_{EW} \sim 10^3$ GeV, which instead is joined
with the particle physics world.  In recent years, models with extra
dimensions have been proposed~\cite{ADD}-\cite{RS1}, where the true
fundamental gravity scale can be as low as a few TeV and where the
Planck mass is a mere effective long-distance 4-dimensional
parameter. For a recent review see~\cite{anton-rev}. In these
scenarios, gravity becomes phenomenologically interesting for high
energy physics and we may have realistic possibilities to observe and
study quantum gravity phenomena at future colliders.

However, one of the constraints on theories with large extra
dimensions is the rate of rare processes such as proton decay.
Current experimental bounds depend on the decay mode in
question \cite{langacker}, with several channels bounded by $\tau_p >
10^{33}$ years \cite{superK,PDG}.  Contributions to 
$p$-decay \cite{perez} include those
from GUT-scale intermediate bosons that mediate baryon number
violation; these contributions can be suppressed below experimental bounds by
additional symmetries as discussed by \cite{ADD, AADD, berezhiani, efn,
kakushadze, bdn}.  In addition, several authors
have worried that proton decay and other rare B- and L-violating decays 
via virtual black holes (BHs) can be
exceedingly rapid. 
According to common belief, the decay/evaporation of
BHs does not conserve any global $U(1)$-quantum numbers and,
in particular, baryonic, B, and leptonic, 
L, charges \cite{stojkovic}. The usual
description of this phenomenon is based on enumeration of all possible
operators which do not conserve global charges, normalized to the
Planck scale with dimensionless coefficients of order unity.

The suggestion that proton decay can be mediated by a virtual 
BH was put forward about 30 years ago by
Zeldovich~\cite{zeld-p-decay} and subsequently considered by
\cite{hpp, page, ehnt, almp}.  The probability of this process in the
standard frameworks with $M_{Pl} \sim 10^{19}$ GeV is very low, but with
smaller Planck mass in TeV range, the baryon number violating
processes through formation of an intermediate virtual BH
become much more efficient.  Adams {\it et al}.
\cite{adams} argued that experimental limits on the proton
lifetime constrain the quantum gravity scale to be larger than
$10^{16}$GeV, implying that the size of the large extra
dimensions is less than $l < 10^{6/n} \times 10^{-30}$cm, where
$n$ is the number of such dimensions.  Arkani-Hamed and Schmaltz
\cite{as} propose reducing the proton decay rate by a scenario
in which the Standard Model fields are confined to a thick wall in
extra dimensions, with fermions ``stuck'' at different points in the
wall; then couplings between them are suppressed due to the
exponentially small overlaps of their wave functions.

Here, instead, we suggest that $p$-decay via virtual BHs may be
more suppressed than previously thought. We propose a ``conjecture''
that, just as in classical gravity, sub-Planck-mass BHs in
quantum gravity can only exist with zero local charge (electric or
color) and zero angular momentum.
In classical general relativity, it can be shown
that a BH with mass smaller than the (effective) Planck mass must have
zero electric charge and zero angular momentum  (see Section 4); 
otherwise no horizon
can exist and the naked singularity is exposed. Moreover, it can be 
shown that it is impossible to make the electric charge of a 
classical BH larger than its mass, $Q^2 > M^2$, in Planck units,
by an influx of charged particles on the horizon (see e.g.~\cite{mtw,lppt}).
In this paper we make
the (as yet unproven) conjecture that this result remains true in
quantum gravity.  Then the virtual BHs that mediate interactions 
such as proton decay must have zero spin and
be electrically neutral. As a
consequence, a four-body collision (3 quarks and a lepton) is required
for formation of the BH intermediate state, which leads to a suppressed
proton decay.

We compute rare baryon and lepton violating decays as well as 
$(n-\bar n)$-oscillations due to intermediate virtual BHs and find that these
rates are in agreement with existing experimental bounds even for TeV
scale fundamental gravity.
In preparation is a second paper, in which we
will investigate a mechanism for generating the baryon number of the
universe using BH-mediated processes.

The content of the paper is as follows.  In Sec. \ref{s-models} 
we briefly review TeV-scale gravity models and in Sec. \ref{s-foam} 
the standard approach to gravitationally induced proton decay. 
In Sec. \ref{s-conjecture} we discuss our BH conjecture.
In Sec. \ref{s-lepton} we consider rare lepton-violating decays 
with a BH intermediate state, giving an estimate of their rates. 
In Sec. \ref{s-baryon}  we discuss 
baryon-violating processes, proton decay and $(n-\bar n)$-oscillations, 
with a BH intermediate state.   
In Sec. \ref{s-heavy} some anomalous decays of particles with heavy
quarks, $t$ and $b$, are evaluated.
The results are summarized in Sec. \ref{s-conclusion}.
{ The paper closes with an appendix where we review
present lower bounds on the magnitude of
the possible fundamental gravity scale $M_*$.}


\section{TeV-gravity models \label{s-models}}

In 1998 Antoniadis, Arkani-Hamed, Dimopoulos and Dvali proposed a
``geometric'' solution to the hierarchy problem of high energy
physics, where the observed weakness of gravity (at long distances)
would be related to the presence of large compact extra
dimensions \cite{ADD,AADD}.  Motivated by string theory, the
observable universe would be a 4-dimensional brane embedded in a
(4+$n$)-dimensional bulk, with the Standard Model particles confined
to the brane, while gravity is allowed to propagate throughout the
bulk.  In such scenarios, the Planck mass $M_{Pl}$ becomes an
effective long-distance 4-dimensional parameter and the relation with
the fundamental gravity scale $M_{\ast}$ is given by
\begin{eqnarray}
M_{Pl}^2\sim M_{\ast}^{2+n}R^n ,
\end{eqnarray}
where $R$ is the size of the extra dimensions.
If these extra dimensions are ``large'', i.e.
$R \gg M_{Pl}^{-1} \sim 10^{-33}$ cm, then the fundamental
gravity scale can be as low as a few TeV and therefore
of the same order of magnitude as $M_{EW}$.
If we assume $M_{\ast} \sim 1$ TeV, we find:
\begin{eqnarray}\label{size}
R \sim 10^{({30}/{n})-17} \: \textrm{cm} .
\end{eqnarray}
In this approach, however, the hierarchy problem is not 
really solved but shifted instead from the hierarchy in energies
to a hierarchy in the size of the extra dimensions which are much
larger than 1/TeV $\sim 10^{-17}$ cm but much smaller than the
4-dimensional universe size. 

The case $n = 1$ is excluded because from Eq. (\ref{size})
we would obtain $R \sim 10^{13}$ cm and therefore strong
deviations from  Newtonian gravity at solar system
distances would result. For $n \ge 2$, $R \lesssim 100$ $\mu$m
and nowadays we have no experimental evidence
against a modification of gravitational forces
in such a regime \cite{r2}. Interesting variations
of these models can lower the fundamental gravity
scale with the use of non-compact extra dimensions
\cite{RS1}.

If gravitational interactions become strong
at the TeV scale, quantum gravity phenomena are
in the accessible range of future experiments
in high energy physics. In particular, there is
a fascinating possibility that hadron
colliders (such as LHC) will be BH factories
(for a review, see e.g. Ref.~\cite{colliders}, criticisms can be
found in Refs.~\cite{mbv,adams}).
From the classical point of view, we expect
BH production in collision
of two particles with center of mass energy
$\sqrt{s}$ if these particles approach each other 
so closely that they happen to be inside the event 
horizon of a BH with mass $M_{BH} \approx \sqrt{s}$.
Semiclassical arguments, valid for
$M_{BH} \gg M_{\ast}$, predict the BH production
cross-section
\begin{eqnarray}
\sigma \approx \pi R^2_{BH}(M_{BH}) ,
\end{eqnarray}
where $R_{BH}(M_{BH})$ is the horizon radius
of a BH of mass $M_{BH}$. 
In the case of an
uncharged and non-rotating BH in (4+$n$)-dimensions,
the horizon radius is obtained from the
(4+$n$)-dimensional Schwarzschild metric \cite{perry}
\footnote{In this paper we use the convention
of Ref. \cite{dimopoulos}:
$ M_{\ast}^{n+2}={1}/{G_{\ast}}$,
where $G_{\ast}$ is the (4+$n$)-dimensional
gravitational constant. See e.g. \cite{giddings}
for other conventions.}
\begin{eqnarray}\label{Schwarzschild}
R_{S} = \frac{1}{\sqrt{\pi}M_{\ast}}
\Big(\frac{M_{BH}}{M_{\ast}}\Big)^{\frac{1}{n+1}}
\Big[\;\frac{8\;
\Gamma(\frac{n+3}{2})}{n+2}\;\Big]^{\frac{1}{n+1}} ,
\end{eqnarray}
where we have ignored possible effects of the
gravitational field of the brane and have
assumed $R_{BH}$ much smaller than the size of
the extra dimensions, so that the boundary
conditions which come from compactification
can be neglected. On the contrary, if $R_{S}$ is larger
than some extra dimensions, the Schwarzschild radius
is given by a lower dimensional BH solution.
We have introduced here two quantities, $R_{BH}$ for the horizon
radius of a generic BH and $R_S$, the same for a  Schwarzschild BH.
Since in what follows we will deal with Schwarzschild BHs only, 
these quantities are the same.


\section{Spacetime foam and proton decay \label{s-foam}}

In this section, we discuss the previous works \cite{almp, adams}
in which BHs from the spacetime foam give rise to proton decay.
We raise some objections and, in the next section,
turn to our conjecture which implies that the decays of this
section do not take place.

Since at the moment a reliable quantum theory of
gravity is lacking, the standard approach is
based on semiclassical calculations. For example, virtual
BHs can be obtained considering the path integral for 
$N$ non-interacting BHs which, in a 4-dimensional
spacetime, is (the generalization to a (4+$n$)-dimensional
spacetime can be found in Ref. \cite{adams})
\be \label{z}
Z \sim \int_0^{\infty} \textrm{d}m
\sum_{N=0}^{\infty} \frac{1}{N!}
\Big(\frac{V}{L^{3}_{Pl}}\Big)^N
\exp \Big(-\frac{4\pi N m^{2}}{M_{Pl}^{2}}\Big) ,
\ee 
where $V$ is a normalization volume and $L_{Pl}$ is
the Planck length. From Eq. (\ref{z}) we can define
a probability distribution of having $N$ BHs with
mass $m$ and, computing the expectation value
of their mass and of their number density, we
find $\langle m \rangle \sim M_{Pl}$ and a density
of roughly one BH per Planck volume. This is the
so called spacetime foam picture \cite{foam1, foam2}
where the spacetime is filled with tiny virtual
BHs which arise as quantum fluctuations out of the 
vacuum. These BHs exist (for very short time, of
order of one Planck time, as a consequence of the
uncertainty principle)
in the same sense 
as electron-positron pairs exist in the 
vacuum of quantum electrodynamics.

In this context we consider the decay of the
proton due to virtual BHs from the spacetime foam. We take
the proton as an object of volume $V_p$
equal to its Compton wavelength cubed, that is
$V_p \sim m_p^{-3}$, with three valence point-like quarks inside.
Probably a better estimate is $V_p \sim \Lambda_{QCD}^{-3}$,
where $\Lambda_{QCD} = 100-300$ MeV. Thus in the estimates presented 
below $\Lambda_{QCD}$ may be substituted instead of $m_p$. 
The probability of finding two quarks in the same Planck
volume is simply given by the ratio of the
Planck volume to the box volume $(m_p/M_{Pl})^3$.
If we include the fact that a virtual BH 
should be formed 
at the same time when these two quarks are packed so closely,
we obtain an additional $m_p/M_{Pl}$ suppression factor
and, by dimensional analysis, we arrive at the decay rate
\be \label{sf-rate}
\Gamma \sim \frac{m_p^5}{M_{Pl}^4} .
\ee
Since BH evaporation conserves energy, charge and angular
momentum, this expression is an estimate of the rate of the
following process:
\be
q + q \rightarrow \bar{q} + l ,
\label{qq-barql}
\ee
where $q$ ($\bar{q}$) is a quark (anti-quark) and $l$
a charged lepton.

The same result can be obtained by different
arguments. The rate of the process (\ref{qq-barql}) can
be estimated as:
\be
\dot n /n = n \sigma_{BH}  = \sigma_{BH} |\psi (0)|^2 ,
\label{dotn-n}
\ee
where $n\sim m_p^3$ is the number density of quarks inside the proton and 
$\sigma_{BH}$ is their interaction cross-section through formation
of a virtual BH. Since the interaction arises from a dimension six operator,
the amplitude has a factor $1/M_{Pl}^2$ and the cross section 
can be estimated as 
\be
\sigma_{BH} \sim m_p^2/M_{Pl}^4 .
\label{sigma-BH}
\ee
Again, we obtain the result in Eq. (\ref{sf-rate}).

Inserting the standard Planck mass $M_{Pl} \sim 10^{19}$ GeV
into Eq. (\ref{sf-rate}) we predict the proton lifetime of the order 
of $10^{45}$ years and there is no problem with the current 
experimental bound.
On the other hand, in models with large extra dimensions
the fundamental gravity scale is $M_*$ and, replacing
$M_{Pl}$ with $M_* \sim 1$ TeV, leads to quite short-lived proton 
with the lifetime 
$\tau_p \sim 10^{-12}$ s. Hence, in order to avoid
contradiction with the present experimental constraints, Adams
{\it et al}. \cite{adams} require $M_* \gtrsim 10^{16}$ GeV. 

However, some criticisms may be raised in this connection. 
In fact, even if in some cases vacuum fluctuations lead
to observable  phenomena, such as the Casimir force, 
they are doubtless not well understood when we
consider quantum gravity effects related to formation
of small virtual BHs.
We may encounter problems, even with
the standard Plank mass of $10^{19}$ GeV,
if we apply the same considerations to leptons, which
are supposed to be elementary ones (not consisting of some
``smaller'' parts). 
In this case much faster gravitationally induced 
decays can be expected, because we do not need to wait 
for two constituent particles to be 
within the same Planck volume. Hence, in the arguments leading to
Eq. (\ref{sf-rate}), only the $m / M_{Pl}$ suppression
factor remains. For example, 
the rate of $\mu\rightarrow e\gamma$ could be
as large as $\Gamma \sim m_\mu^2 / M_{Pl}$. This estimate
is based on time uncertainty relation and gives rise to the
decay branching ratio about $10^{-4}$, which contradicts
the experimental bound by 6 orders of magnitude.

However, these considerations are pretty inaccurate
and cannot rigorously lead us to the 
conclusion that the spacetime foam picture with the
standard Planck mass is incorrect. In
particular, the result presented above for the rate of the
decay $\mu\rightarrow e\gamma$, which is inversely proportional to
the first power of the Planck mass looks very suspicious, one
should expect at least the second power, and possibly indicates 
a sickness of this kind of argument, because in any normal
theory the probability of a process in the leading order
must contain an even power of the coupling constant.
This is true in the usual approach, according to which the decay 
$\mu\rightarrow e\gamma$ is described by the dimension five operator,
$F_{\mu\nu} \langle \psi \sigma^{\mu\nu} \psi \rangle/M_{Pl}$
so that $\Gamma \sim m_\mu^3/M_{pl}^2$. It
does not contradict the experimental bound with the normal Planck 
mass, $M_{Pl} = 10^{19}$ GeV, but still it puts a very strong lower 
limit on the effective Planck mass, $M_* \gtrsim 10^{13}$ GeV.

In fact practically all ``natural'' estimates with
the low scale gravity lead to too high rates of processes with
baryonic and/or leptonic number violation. One possible conclusion
is that the gravity scale must be high, close to $M_{Pl}$, or
that there is some new exclusion principle which forbids an
easy formation of virtual BH. In what follows we will explore
the latter option, requiring zero charge and angular 
momentum of virtual BHs because non-zero charge and
momentum are classically forbidden for very light ones 
with $m<M_*$. Simultaneously here enters another 
important assumption that
virtual BHs have masses of the order of the masses of the initial 
state (see the next section).

\section{A Classical Black Hole Conjecture \label{s-conjecture}}

In classical general relativity, a BH with mass smaller than the (effective)
Planck mass cannot be formed if it has a non-zero electric 
charge or if it rotates.  Here we will make the conjecture that
this result remains true in quantum gravity.  Hence the virtual
BHs that mediate interactions such as proton decay
must be electrically neutral and spinless. 

The classical condition that a
BH less massive than the Planck mass
must have vanishing angular momentum and
electric charge can be seen as follows. We can examine the
4 dimensional Kerr-Neumann or Reissner-N{\"o}rdstrom solutions 
to see that a horizon does not exist if the BHs have
electric charge or angular momentum.  The position of the horizon
is given by the expression~\cite{mtw,lppt}
\be
R_{BH} M_{Pl} = \frac{M_{BH}}{M_{Pl}} + 
\sqrt{\left( \frac{M_{BH}}{M_{Pl}}\right)^2 - Q^2 - J^2  
\left(\frac{M_{BH}}{M_{Pl}}\right)^{-2}} ,
\label{r-hor}
\ee 
where $Q$ and $J$ are respectively electric charge and angular
momentum of BH. It is clear that in 4D space-time the
horizon cannot be formed if
\be
\left( \frac{M_{BH}}{M_{Pl}}\right)^2 < \frac{Q^2}{2} +
\sqrt{ \frac{Q^4}{4} + J^2} .
\label{mass-limit}
\ee
In the absence of a horizon, there would be naked singularities
which are not allowed.
Indeed if condition (\ref{mass-limit}) is fulfilled,
the Kerr-Newmann metric allows to reach the 
physical singular point 
$r=0$ from some large $r$ in finite time
without crossing any metric singularity.
Probably BHs with such low masses cannot be created.
See e.g. problem 17.17 in book~\cite{lppt} where it is
shown that it is impossible to charge an already existing 
Reissner-N{\"o}rdstrom BH up to $Q>M_{BH}/M_{Pl}$
by any physical process. So a naked singularity cannot be created. 

This result is true only in classical physics and quantum gravity effects 
may change the bound (\ref{mass-limit}). 
Moreover, the time-energy uncertainty
relation could allow a virtual BH to have mass/energy larger
than the mass of the initial particle. Still in the absence of anything 
better we assume that only diagrams with uncharged and non-rotating
virtual BHs can participate in light ($m<M_{Pl}$) particle decays.

As for the color charge, we expect a similar picture, though the 
situation is much less clear. 
For classically large BHs any color charge
should be screened at the microscopic distances of the order of the 
inverse QCD scale, $R_{QCD} = 1/\Lambda_{QCD}$, so there are no 
colored hairs at large distances but they can be present at the distances
smaller than $R_{QCD}$. The analysis of classical
colored BH properties can be found in Ref.~\cite{bizon}.
For microscopically small BH, considered here,
the screening radius $R_{QCD}$ is usually large in comparison with
the radius of the BH. Thus for sub-Planck-mass BHs we can 
apply the same prescription as is done above for electrically 
charged BH and require colorless states, even if an analytic bound 
such as Eq. (\ref{mass-limit}) does not exist. 
On the other hand, weak $SU(2)$ charges can be safely neglected, 
since the symmetry is broken via the Higgs mechanism
and BHs do not manifest charges if they are related to
massive gauge bosons \cite{dolgovj}. To be more precise, weak
hairs may exist but only for a short time, $\tau_W \sim 1/M_W$,
where $M_W$ is the mass of weak gauge boson, $W$ or $Z$. Since this 
time is much shorter than the life-time of the processes
considered below we neglect weak hairs of BHs and allow for arbitrary
quantum numbers with respect to weak $SU(2)$.

A similar situation should hold in higher dimensions but instead 
of $M_{Pl}$ an effective gravitational scale $M_*$ should be
substituted. If the Compton wavelength of an
elementary particle, $\lambda_C = 1/m$, is much smaller than
the size of the extra dimensions, it is natural to believe 
that gravity ``inside'' 
an elementary particle becomes multidimensional.

In addition to this conjecture of neutral and non-rotating
BHs we impose some, maybe even more questionable, conditions 
in calculations/estimates of the amplitude of reactions
with broken global quantum numbers due to virtual BH. 
In essence we suggest a set of rules which do not 
respect some of the usual conditions existing in quantum field theory, in
particular crossing relations between amplitudes. For example,
we allow a virtual BH to decay into, say, a proton and a electron,
but we do not allow a proton to form a BH plus a positron,
with the same amplitude. The picture that we have in mind is a kind 
of time ordering: a BH could be formed in a collision of a neutral system
of particles in the s-channel whereas a BH 
cannot be in the t-channel of a reaction.
We assume that BHs can be formed out of positive energies of real
particles only and not from virtual energies of particles in closed loops.
For example, BH cannot be formed by vacuum fluctuations, despite the fact
that, according
to the standard picture, vacuum fluctuations might create a pair or more of
virtual particles both with positive and negative energies.
The mass of the 
BH should be of the order of the energy of incoming (or outgoing)
particles.
In an attempt to describe this in terms of the usual language we come to 
a version of the old non-covariant perturbation theory with all virtual 
particles having positive energies. It corresponds to the choice of only one
mass-shell pole in the Feynman Green's functions. This rule allows
only for BHs with masses which are of the order of the energies of the
initial (or final) particles, as we postulated above.
It may look very strange, to say the least, but virtual BHs are not
well defined objects and we do not know what happens with space-time
at the relevant scales. Taken literally these rules would lead to
violation of some sacred principles of the standard theory (locality, Lorentz 
invariance, and more). Let us remind the reader, however, that the
existing attempts in the literature to invoke virtual BHs are based
on standard quantum field theory in a situation where it is 
almost surely inapplicable.

So it is not excluded that many properties
of the standard field theory are broken, including even Lorentz 
invariance and locality. We cannot of course present any serious
arguments in favor of our construction but it predicts quite
impressive phenomena with clear signatures based on a
very simple set of rules and if these 
effects are discovered, the approach may be taken more seriously.
Our goal here is to formulate a reasonable(?) set of rules
which may possibly describe processes with virtual BHs and 
are, at least, not self-contradictory. Based on these rules we will
study the phenomenological consequences which are quite rich and
may be accessible to experiments after a minor increase of accuracy.


\section{Lepton number violating decays mediated by black holes 
\label{s-lepton}}

Gravitational interactions are known to become
stronger at short distances or at high energies
and in the standard theory they are expected to become strong at 
$E\sim M_{Pl} \sim 10^{19}$ GeV or at the distances about
$10^{-33}$ cm. In TeV-gravity models,
we can expect non-negligible gravitational effects
much earlier at the energy scales of contemporary
particle physics, at relatively low energies, about electroweak
scale. Let us remind the reader that rare decays have been
often used to probe small distances and
sometimes have given information about heavy
particles prior to their discovery. They can
play the same important role to
investigate models with the fundamental gravity
scale in the TeV range.

\subsection{Muon decay $\mu \rightarrow 3e$ \label{ss-mu3e}}

In this connection we will consider $\mu^-$ decay
into $e^-e^-e^+$, which violates muonic/electronic number
conservation due to formation of a virtual BH in the
intermediate state. The diagram describing this process is
presented in Figure \ref{f-muon}$a$.

First, the muon emits a virtual photon; then the photon produces an
electron-positron pair. Next the muon and the positron form a
Schwarzschild BH.  Since the BH does not
respect muon-number conservation, it can decay into an $e^+e^-$-pair.
This is not the Hawking radiation \cite{hawking} because the latter is a
semiclassical process, which can be realized for ``large'' BHs, 
but is essentially a quantum gravity phenomenon. We cannot
reliably calculate its amplitude, since it is surely non-perturbative,  
but assume that it can be estimated
on dimensional grounds assuming that numerical coefficients are of
order unity.  We also assume that the BH decays predominantly into
Standard Model particles on our brane \cite{horowitz}
(however, since bulk emission of gravitons becomes more relevant
as the number of extra dimensions increases \cite{cardoso}, 
even processes with missing energy may play an interesting role). 
BH decay conserves energy, angular momentum and gauge charges, but violates
global symmetries. As we have already mentioned the process presented
in Figure \ref{f-muon}$a$ violates the conservation of lepton family number.

Let us now make a rough estimate of the
rate of this decay. The emission of the virtual
photon and its subsequent transformation
into an $e^+e^-$-pair leads to the suppression of the
probability by the factor of the
order of $(\alpha/2\pi)^2$. 

By dimensional arguments, the amplitude of the 
decay
$\mu^-e^+$ into $e^+e^-$ through a virtual BH is 
proportional to $g_2^2/M_{BH}^2$, where $M_{BH}$ is the mass of BH 
and $g_2$ is the dimensionless coupling constant of BH to two fermions. 
This coupling constant must be proportional to the strength of
gravitational coupling. Here we make the assumption that
$g_2 \sim R_S E$, where
$E$ is the total energy of all colliding particles which
make the virtual BH in their center of mass system, i.e. 
$E=M_{BH}$. Thus $g_2/M_{BH} = R_S$. As is written above,
$R_S$ is the Schwarzschild gravitational radius.
For $R_S$ we use Eq. (\ref{Schwarzschild}) with the last 
factor in square brackets taken to be $\sqrt{\pi}$ in the case of
multidimensional gravity, while $R_S = M_{BH}/M_{Pl}^2$ for the
standard gravity.

We have identified the energy of the colliding or outgoing particles
with the BH mass, while $R_S$ describes the
strength (or better to say, weakness) of gravitational interaction.
We emphasize that this is an assumption about the nature of the
(quantum) gravitational interaction with the BH.
As for the mass of the BH we assume that it is of the order of
the muon mass, $M_{BH}\approx m_\mu$, or more precisely,
$M_{BH} = E_{e^+}+E_{e^-} $ in their rest frame.
In principle one might worry that a
virtual BH might have a much larger mass, e.g. larger than $M_*$, 
and in this case it could be both charged and rotating. For very heavy 
BHs in the intermediate state the amplitude should be suppressed 
by an inverse power of BH mass and hence it could be weaker than
the amplitude with light neutral BH. On the other hand, if a heavy
charged and/or rotating BH could be formed in low-body (e.g. two-body)
collision of the constituents of the initial particle, while neutral
BH demands more particles for its creation, the processes with
heavy BHs would be less suppressed and could be dangerously efficient,
despite an absence of the factor $(\alpha/2\pi)^2$. A much larger
$M_*$, beyond TeV range would then be necessary in this case to avoid conflict 
with experiment.  

Quantum field theory allows for any masses of virtual particles. However,
it may be not true for virtual BHs which are surely nonperturbative 
and quite complicated quantum fluctuations of space-time. We suggest
that such heavy intermediate black holes are not allowed.
Our hypothesis is that virtual BH's are in some sense real, i.e.
their mass is of the order of the real characteristic energy of
colliding particles which form BH in the process under scrutiny. 
This assumption contradicts the statement of quantization of 
BH masses with steps of the order of $M_*$ but  in the absence of 
a serious theory it may be as good as any other. At least, it is very 
simple and in some sense natural: for formation of BH we need real
available energy which is of the order of the decaying particle
mass.

The amplitude of $\mu\rar 3e$ decay corresponding to the diagram in
Fig. 1$a$ is equal to:
\be
A(\mu\rar 3e) = \frac{\alpha\,\ln \left({M_*^2}/{m_\mu^2}\right)}
{\pi\, M_*^2}\,
\left[ \left(\gamma_{23}\right)^{\frac{2}{n+1}}
(\bar\psi_2 \psi_3)(\bar\psi_1 \psi_\mu) -
 \left(\gamma_{13}\right)^{\frac{2}{n+1}}
(\bar\psi_1 \psi_3)(\bar\psi_2 \psi_\mu) \right] ,
\label{a-mu-3e}
\ee
where $\gamma_{ij} =  M_{BH}^{(ij)}/M_*$ come from the factor
$g_2^2/M_{BH}^2=R_S^2$ (see Eq. (\ref{Schwarzschild})) and
$\alpha =1/137$; the log-factor comes from logarithmically 
divergent triangle part of the diagram with the ultraviolet
cut-off taken at the gravity scale $\Lambda_{UV}\sim M_*$; 
$\psi_j$ are the Dirac spinor wave functions of the corresponding
particles. The amplitude is antisymmetric with respect to interchange of 
two final state electrons 1 and 2. The mass of the virtual BH is taken
to be equal to the energy of the $e^+e^-$ pair emitted by this BH. The
upper indices $M_{BH}^{(ij)}$ indicate which particles are emitted
by BH. In what follows we substitute for simplicity
for $M_{BH}^{(ij)}$ its average value
\be
\langle \left(M_{BH}^{(12)}\right)^2 \rangle = 
\langle \left( m_\mu^2-2m_\mu E_3\right)  \rangle = \kappa m_\mu^2 ,
\label{m-bh-av}
\ee
where $\kappa = 1/2\, -\, 1/3$.
Taking the necessary traces and integrating over three body phase space
we find the decay width:
\be
\Gamma(\mu\rar 3e)_n = \frac{\alpha^2 m_\mu}{2^{11}\pi^5}\,
\left(\ln \frac{M_*^2}{m_\mu^2}\right)^2\,
\left(\frac{m_\mu}{M_*}\right)^{4(1+\frac{1}{n+1})}
\kappa^{\frac{2}{n+1}} .
\label{gamma-mu-3e}
\ee
Two factors 1/2 come from averaging over the spin of the initial muon and 
from 1/2! due to two identical particles in the final state.

In the case of the ordinary (3+1)-dimensional gravity with $M_* =M_{Pl}$,
the decay rate is roughly equal to
\be
\Gamma (\mu\rar 3e)_{4D} \sim
0.001\left(\frac{\alpha}{2\pi}\right)^2\, \frac{m_\mu^9}{M_{Pl}^8}
\label{mu-3e-4dim}
\ee 
and is negligibly small. In the higher
dimensional case with $n=2$:
\be
\Gamma (\mu\rar 3e)_{2} \approx 6 \cdot 10^{-31}\,{\rm GeV}\,
\left( 1 + 0.11\ln \frac{M_*}{ {\rm TeV}} \right)^2 \,
\left(\frac{{\rm TeV}}{M_*}\right)^{\frac{16}{3}}\,
\left( 3\kappa\right)^{2/3}.
\label{gamma-num}
\ee

Since the total decay rate of the muon is
\begin{eqnarray}
\Gamma_{tot} \approx 3 \cdot 10^{-19}
\: \textrm{GeV}
\label{gamma-mu-tot}
\end{eqnarray}
then, assuming $M_{\ast} \sim 1$ TeV, 
we obtain the branching ratio about $2\cdot 10^{-12}$ for $n=2$
which is close to the present experimental 
constraint~\cite{PDG}:
\begin{eqnarray}
BR(\mu^- \rightarrow e^-e^+e^-)\,\Big|_{Exp}
< 1.0 \cdot 10^{-12} .
\label{br-mu-3e}
\end{eqnarray} 
For larger $n$ the decay rate would be in stronger disagreement with the
experimental bound. However even a minor increase in the value 
of $M_*$ would
avoid the contradiction. It is intriguing that these estimates 
with $M_* \sim $ TeV are quite close to the existing bounds on
muon number violating decay $\mu\rar 3e$.

\subsection{Lepton number violation in $e^+e^-$ collisions \label{ss-eemue}}

The conservation of muon number can be also violated in the reaction
\be
e^+ + e^- \rar \mu+ e
\label{ee-mue}
\ee
or similar reactions with any other leptons in the final state. A virtual
BH can be formed here just from the initial $e^+e^-$ pair and the 
cross-section of this reaction would be about 
\be
\sigma(e^+e^- \rar \mu e) \approx 7\cdot 10^{-39} \,{\rm cm}^2\,
\left(\frac{M_{BH}}{100\,{\rm GeV}}\right)^{2+\frac{4}{n+1}}\,
\left(\frac{{\rm TeV}}{M_{*}}\right)^{4+\frac{4}{n+1}} .
\label{sigma-e-mu}
\ee 
This may be observable in high energy $e^+e^-$ collisions.
We again note that the predicted cross-section (\ref{sigma-e-mu})
is surprisingly close to the current limits on lepton flavor
violation search at the $Z^0$ resonance \cite{opal}.
One should keep in mind, however, that the total angular momentum 
of the initial $e^+e^-$ pair must vanish. This means that the annihilation
should proceed from the
$s$-state because the total spin of the initial particles is 
zero. This demands the same sign of helicity of $e^+$ and $e^-$ in their
center of mass frame. On the other hand, the annihilation through $Z$-boson
proceeds with total angular momentum equal one and demands opposite helicities
of $e^+$ and $e^-$.

\subsection{Muon decay $\mu \rightarrow e \gamma$ \label{ss-muegamma}}

One can obtain an estimate of the decay rate $\mu\rightarrow e\gamma$
along the same lines. In the simplest diagram (see Figure 1$b$)
the probability is suppressed by an additional
power of $\alpha$, but the ratio of
two-body to three-body phase space compensates this extra suppression,
so that the probability of $\mu\rightarrow e\gamma$ decay through the
considered mechanism would be approximately the same as
$\mu\rightarrow 3e$. 

We should keep in mind however, that there is an additional ambiguity
related to the second loop in diagram of 
Fig.~\ref{f-muon}$b$, which contains three virtual
electrons. If we first integrate over the $(\mu\,e\,\gamma)$-loop, the second
loop would also be logarithmic, because one electron propagator, common 
for both loops, would be reduced to a point. The log-factor related to
the second loop should be taken with the loop counting factor, $1/(2\pi)^2$
and the result would be somewhat smaller than the naive estimate without 
logarithmic and loop counting factors. 
There is an important difference between the first and the second loops.
The particles in the first loop interact with the BH at one point and 
the BH propagator
does not enter into consideration. In the second loop the interaction with 
the BH
is not pointlike and the statement about its ultraviolet behavior was made
assuming the standard form of the 
BH propagator, $\sim (P^2_{BH} - M_{BH}^2)^{-1}$.
It is unknown whether or not this is true.  
Possibly covariant perturbation theory
is not applicable to the case of virtual BH and naive estimates of divergences
are not valid (see Sec.~\ref{s-conjecture}).
This problem is also encountered 
in Sec. \ref{ss-susy},
where it is essential for the calculation of the time of 
$(n-\bar n)$-oscillations.

\subsection{$\tau$ decays \label{ss-tau}}

The widths of the 
analogous processes with nonconservation of the $\tau$-flavor number
in $\tau$-lepton decays would be enhanced by the ratio 
$(m_\tau/m_\mu)^{5+4/(n+1)}$, which is  $6\cdot 10^7$ for $n=2$. 
Since the $\tau$ life-time is $10^7$ times shorter than that of the muon, 
the expected branching
ratios of the decays $\tau\rar 3l$ or $\tau\rar l\gamma$ would be an order of
magnitude larger than those for similar muon decays, i.e. maximally 
around $10^{-11}$.  It surely does not
contradict the existing bounds, which are of order of
$10^{-6}-10^{-7}$~\cite{PDG}. It is unclear if such decays may be reachable
in not too distant future.

If the virtual BH in the diagram in Fig. \ref{f-muon}$a$ 
emits a $q\bar q$-pair, instead of 
a lepton-antilepton pair, it would lead to decay of $\tau$-lepton into
semileptonic channel, $\tau^-\rar l^- +$ mesons. The branching
ratio (BR) for this inclusive process is expected at the same level as
the 
BR for $\tau\rar 3e$ decays, or somewhat smaller, because a decay of BH into
a pseudoscalar state (e.g. $\pi$-meson) may be suppressed and the dominant
decay channel should be a scalar one.

There can also be interesting modes of $\tau$ decays with 
non-conservation of baryon and lepton numbers, as e.g.
$\tau^- \rar e^-e^+ \bar p$, $e^-e^-p$ and analogous ones with
neutrons and neutrinos. The diagrams describing such decays are essentially
the same as Fig. \ref{f-muon}$a$, 
with replacement of the initial $\mu$ with $\tau$ and with a BH which
emits three quarks and one lepton instead of two leptons.
The effective Lagrangian describing e.g. such decays has the  
form:
\be
\frac{\alpha}{\pi}\,\ln \left({M_*^2}/{m_\tau^2}\right)
\frac{g_2 g_4}{M_{BH}^2}\,\bar\psi_l \psi_\tau \psi_q^3 \psi_l ,
\label{L-tau-3qll}
\ee
where different $\psi$'s are spinor wave 
functions/operator of the corresponding
fermions and $g_n$ is the coupling constant of 
BH to $n$ fermions. According to the
arguments presented in the preceding sections, creation of BHs in multiparticle
collisions should be suppressed due to the necessity for several
particles to meet in the same small volume
and thus, e.g. $g_4$ should contain the BH radius to the fourth power,
while the rest of the necessary dimension is supplied by the BH mass:
\be
g_4 = R_S^4 M_{BH},\,\,\,{\rm and}\,\,\, g_2 = R_S M_{BH} .
\label{g4-g2}
\ee
The amplitude of the decay $\tau \rar p l \bar l$ is determined by the matrix
element of this operator between an initial $\tau$ and final $3l+p$ states.
The matrix element of making a proton out of three quarks is:
\be
\langle p | \psi_q^3 | {\rm vac} 
\rangle \sim \frac{m_q^3}{(2\pi)^4} \bar u_p ,
\label{p-3q}
\ee
where $m_q \approx 300$ MeV is the constituent quark mass, 
that is the characteristic energy scale of the process,
and $u_p$ is the spinor wave function of the proton.

We do not distinguish in this section between the masses of 
light constituent quarks, $m_q \sim 300$ MeV and the characteristic scale of 
strong interaction, $\Lambda_{QCD} \sim 100$ MeV. 
However, in future sections of the paper, where we  
consider proton decay and
neutron-antineutron oscillations, the difference may be important.

It is straightforward now to calculate the decay rate 
\be
\Gamma(\tau \rar pl\bar l) \sim 10^{-3} \left(\frac{\alpha}{2\pi}\right)^2\,
\left(\ln \frac{M_*^2}{m_\tau^2}\right)^2 \,
\left(\frac{m_q}{M_*}\right)^6 \,
\left(\frac{m_\tau}{M_*}\right)^{4 + \frac{10}{n+1}} m_{\tau} 
\label{gamma-tau-3q}
\ee
where the first factor, $10^{-3}$, comes from three-body phase space.
The life-time with respect to this decay would be extremely long:
for $M_* \sim 1$ TeV and $n = 2$, we obtain $\tau \sim 10^{16}$ years. 
We have omitted here the loop counting factors but the effect is tiny
even without them.

\subsection{$K$-mesons decays \label{ss-kaon}}

Similar considerations can be applied to rare decays
of $K$-mesons. In particular, if two quarks constituting $K^0$-meson
might form BH, this BH could decay into any neutral combination of
two leptons, $e^+e^-, \mu^+\mu^-$ and $\mu^\pm e^\mp$. 
The amplitude of this decay can be estimated as:
\be
A(K\rar 2l) \approx g_{Kqq}\,\left( \frac{g_2}{M_{BH}} \right)^2 \,
\frac{m^2_q}{(2\pi)^2}\, \bar\psi_1 \psi_2 ,
\label{K-2l}
\ee
where $g_{Kqq}$ is the Yukawa coupling constant of $K$-meson to $\bar q q$-pair
and the third factor came from the integration of the loop
of $\bar q q$ pair which combines into $K$-meson. 
The corresponding decay width is given by
\be
\Gamma(K^0 \rar l\bar l) 
= \frac{g_{Kqq}^2 m_K}{4\,(2\pi)^5}\,\left(\frac{m_q}{M_*}\right)^4\,
\left(\frac{m_K}{M_*}\right)^{\frac{4}{n+1}} .
\label{k2l}
\ee 
We have assumed that $M_{BH}= m_K$
and $E=m_K$ since no energy is taken away to other particles.
With $g_{Kqq} \sim 1$,
$m_q = 300$ MeV, $M_* = 1$ TeV and $n=2$, we obtain for
the life-time of this decay 
\be
\tau(K^0 \rar l\bar l) \approx 0.16\,\,{\rm s} .
\label{tau-K0-ll}
\ee
For $K^0_2$ it gives the branching ratio 
$BR(K^0 \rar l\bar l) \sim 10^{-7} $, which is 4-5 orders of magnitude
above the experimental bounds: for example, 
$BR(K^0 \rar e^+e^-) < 10^{-11} $ and 
$BR(K^0 \rar e^\pm \mu^\mp)  <5\cdot 10^{-12} $. 
To avoid the contradiction
we should either take $M_* > 3$ TeV or to see if there is a possible
suppression mechanism for this decay in this scenario.
In fact, there is one. It is natural to expect that BH should have
the quantum numbers of the vacuum, i.e. it should be a scalar object.
Hence the $K$-meson, which is a pseudoscalar, cannot transform to BH directly,
but should emit some other particle in such a way that the remaining 
combination of the quark-antiquark system would be scalar.
The simplest way is to emit a $\pi^0$-meson,
while the remainder would make a BH which would decay into $l\bar l$.
This mechanism means in particular that two body decays 
$K\rar l\bar l$ are suppressed and not dangerous.

The amplitude of the decay $K\rar \pi ll$ is described by the diagram
in Fig. \ref{f-kaon} 
and is equal to 
\be
A(K\rar \pi ll ) = \frac{g^2_2\, g_{K\pi S}\, m_q^2}{(2\pi)^2 M_{BH}^2}\,
\bar\psi_{l_1} \psi_{l_2}
\label{a-k-pi-ll}
\ee
where $g_{K\pi S} $, the coupling constant of $K$ and $\pi$ to the scalar
state of quark-antiquark pair, has dimension of inverse mass. The factor
$m_q^2/(2\pi)^2$ comes from the quark loop.
The mass
of BH is $M^2_{BH} = (p_{l_1} + p_{l_2})^2 = m_K^2 + m_\pi^2 - 2 m_K E_\pi $.
The life-time with respect to this decay, after integration over 3-body
phase space with $M_{BH}$ depending upon the pion energy, $E_\pi$, 
is equal to:
\be
\tau(K\rar \pi ll ) &=& 0.85 \cdot 10^2\,{\rm s}\, 
\left(g_{K\pi S} m_\pi \right)^{-2}\,
\left(\frac{M_*}{{\rm TeV}}\right)^{4 + \frac{4}{n+1}}
\left(\frac{{\rm TeV}}{m_K}\right)^{\frac{4}{n+1}-\frac{4}{3}}
\cdot \nonumber\\ && \quad \cdot
\left(\frac{300\,{\rm MeV}}{m_q}\right)^4 \,
\frac{6.4 \cdot 10^{-3}}{f_n},
\label{tau-k-pi-ll}
\ee
where $f_n$ is related to integration over phase space:
\be
f_n = \int_{\mu}^{(1+\mu^2)/2} dx\,\sqrt{x^2-\mu^2}\,
\left(1+\mu^2 -2x\right)^{1+\frac{2}{n+1}}.
\label{fkn}
\ee 
Here $\mu = m_\pi/m_K$. The factor $6.4\cdot 10^{-3}/f_n$
is equal to
1 for $n=2$, to 0.82 for $n=3$, and to 0.58 for $n=7$.   

The experimental bounds on the branching ratios of $K_L^0\rar \pi^0 2l$ 
decays are about $(3-5)\cdot 10^{-10}$, while its life-time is 
$5.2\cdot 10^{-8}$ sec. Thus for $n=2$, $M_* = 1 $ TeV, and
$g_{K\pi S} m_\pi \sim 1$ we are
on the verge of discovery of such decays. For larger $n$ a larger $M_*$
or smaller $g_{K\pi S}$ are needed.

Similar estimates can be presented for decays of charged $K$-mesons,
e.g. to
\begin{eqnarray}
K^+ \rightarrow \pi^+\,\mu^+\, e^- ,\,\,\,\pi^+\,2\nu, \,\,\, etc .
\label{k-pi-mu-e}
\end{eqnarray}
Experimental bounds on the branching ratios of these decays are 
about $3\cdot 10^{-11}$~\cite{PDG}. The absolute probability
of decay (\ref{k-pi-mu-e}) is approximately the same as that
of $K^0\rar \pi^0 l \bar l$. Since the total life-time of $K^+$
is 4 times shorter than that of $K_2^0$, the predicted
branching ratios are 4 times smaller and, according to the 
discussion above, they may be close to the existing bounds.

This model has a few interesting features/signatures. The dominant 
anomalous decay mode is three body. The charge of the emitted pion is
the same as the charge of the initial $K$. The probabilities of the
decays with charged and neutral leptons in the final states are
approximately the same. The rather large magnitude of the branching
ratios of these anomalous decays of $K$-mesons
make them very interesting/promising candidates in the search
for non-conservation of global lepton quantum numbers.

Moreover, we note another appealing feature of all the decays with
a BH intermediate state: if undiscovered weakly interactive
light particles exist, such as for example possible sterile
neutrinos or axions, they should be emitted by the BH with the
same probability of the other light particles,
if compatible with the BH quantum numbers. In fact, tiny
BHs may provide a unique opportunity to discover such a
kind of particles because, if the so-called ``equivalence principle''
holds, gravity couples to any form of energy with the
same strength and does not distinguish one type of
particle from another. Obviously, the emitted weakly
interactive particles cannot be seen by the detector
and the event appears as a process with violation
of energy, momentum and, possibly, angular momentum.

\begin{center}
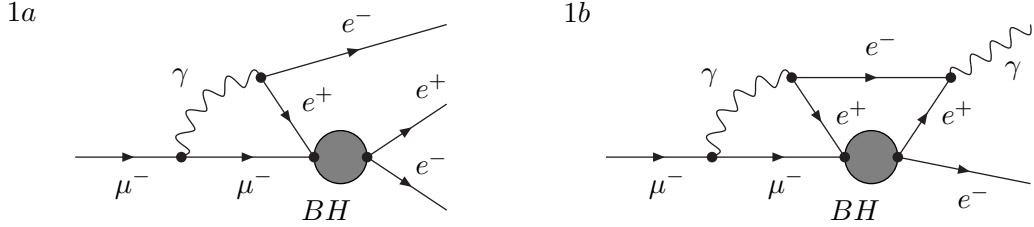
\begin{figure}
\begin{picture}(340,85)(0,0)
\Text(20,80)[]{$1a$}
\ArrowLine(40,25)(80,25)
\ArrowLine(80,25)(130,25)
\PhotonArc(110,25)(30,90,180){3}{5}
\ArrowLine(110,55)(180,75)
\ArrowLine(110,55)(130,25)
\ArrowLine(150,25)(180,45)
\ArrowLine(150,25)(180,5)
\GCirc(140,25){10}{0.5}
\Text(62,15)[]{$\mu^-$}
\Text(108,15)[]{$\mu^-$}
\Text(80,55)[]{$\gamma$}
\Text(132,47)[]{$e^+$}
\Vertex(80,25){2}
\Vertex(110,55){2}
\Vertex(130,25){2}
\Vertex(150,25){2}
\Text(135,5)[]{$BH$}
\Text(147,77)[]{$e^-$}
\Text(175,52)[]{$e^+$}
\Text(175,23)[]{$e^-$}
\Text(230,80)[]{$1b$}
\ArrowLine(240,25)(280,25)
\ArrowLine(280,25)(330,25)
\PhotonArc(310,25)(30,90,180){3}{5}
\ArrowLine(310,55)(370,55)
\ArrowLine(310,55)(330,25)
\ArrowLine(350,25)(370,55)
\ArrowLine(350,25)(400,15)
\Photon(370,55)(400,75){3}{4}
\GCirc(340,25){10}{0.5}
\Text(264,15)[]{$\mu^-$}
\Text(309,15)[]{$\mu^-$}
\Text(280,55)[]{$\gamma$}
\Text(334,42)[]{$e^+$}
\Vertex(280,25){2}
\Vertex(310,55){2}
\Vertex(330,25){2}
\Vertex(350,25){2}
\Vertex(370,55){2}
\Text(335,5)[]{$BH$}
\Text(345,68)[]{$e^-$}
\Text(374,42)[]{$e^+$}
\Text(380,10)[]{$e^-$}
\Text(395,58)[]{$\gamma$}
\end{picture}
\caption{$a)$ Muon decay $\mu \rightarrow 3e$
with BH intermediate state.
 $b)$ Diagram for
the process $\mu \rightarrow e \gamma$.}
\label{f-muon}
\end{figure}
\end{center}

\begin{center}
\begin{figure}
\begin{picture}(340,110)(0,0)
\ArrowLine(100,31)(151,31)
\ArrowLine(100,26)(151,26)
\ArrowLine(151,30.5)(211,90.5)
\ArrowLine(155,26.5)(215,86.5)
\Line(155,31)(175,31)
\CArc(175,36)(5,270,360)
\CArc(185,36)(5,90,180)
\Line(155,26)(175,26)
\CArc(175,21)(5,0,90)
\CArc(185,21)(5,180,270)
\ArrowLine(185,41)(225,41)
\ArrowLine(185,16)(225,16)
\CArc(225,6)(35,40,90)
\CArc(225,51)(35,270,320)
\ArrowLine(271,28.5)(321,48.5)
\ArrowLine(271,28.5)(321,8.5)
\GCirc(261,28.5){10}{0.5}
\Text(257,8.5)[]{$BH$}
\Text(215,50)[]{$q$}
\Text(215,5)[]{${\bar q}$}
\Text(215,30)[]{${\rm J^P} = 0^+$}
\Text(230,100)[]{$\pi^0$}
\Text(315,55.5)[]{$e^-$}
\Text(315,26.5)[]{$\mu^+$}
\Text(120,45)[]{$K^0$}
\GCirc(153,28.5){6}{0.8}
\Vertex(251,28.5){2}
\Vertex(271,28.5){2}
\end{picture}
\caption{Kaon decay $K^0 \rightarrow \pi^0 e^- \mu^+$.
$K$-meson probably cannot transform directly to a BH,
because it is a pseudoscalar particle; instead, the emission
of a $\pi^0$ leaves a scalar $q \bar q$ system.}
\label{f-kaon}
\end{figure}
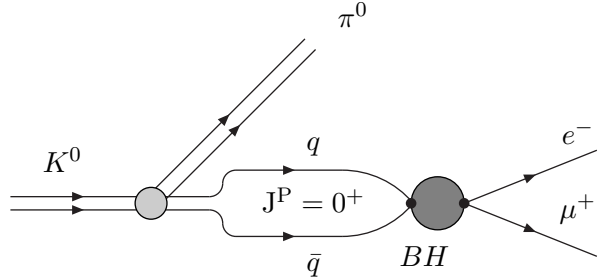
\end{center}

\section{Baryon number violating processes mediated by black holes
\label{s-baryon}}

\subsection{Proton decays \label{ss-proton}}

Proton decay is described by a more complicated diagram, as shown in
Fig.~\ref{f-proton}. 
To create an electrically neutral, colorless,
nonrotating BH, one of the three quarks in the proton must emit a
lepton pair through $\gamma$, $W$, or $Z$ exchange; one of these
leptons, together with the three quarks, may then form such a BH devoid
of any quantum numbers.

Since the BH is now formed in a 4-body collision, 
the probability of BH creation would be suppressed by an additional
small ratio squared of BH volume to the proton volume.
The ratio of volumes appears in the second power because we 
have two more particles in the initial state in comparison with
BH creation in $\mu \rar 3e$ decay. In the latter case the BH could be 
produced in a two-body collision.
This leads to the suppression factor
$(R_S \Lambda_{QCD} )^6$, where $\Lambda_{QCD} \sim 100$ MeV 
is the inverse proton size (in what follows we skip the sub-QCD). 
In fact, the situation
is more complicated and a one parameter description may be impossible
because the proton size is roughly the inverse pion mass, while the
characteristic quark energies in the proton, which determine the
coupling to BH are of the order of the constituent 
quark mass, i.e. $m_q \approx 300$ MeV.  

We will come to the same conclusion studying the transformation
$3q e^- \rar \bar q q$ through a virtual BH, as is 
considered in Sec. \ref{s-lepton}. The amplitude of the reaction
is proportional to $g_2\, g_4/M^2_{BH}= R_S^5 $, where $g_n$ is the 
coupling constant of BH to $n$ fermions; $g_2$ is dimensionless, while
$g_4$ has dimension of inverse mass cubed, see Eq. (\ref{g4-g2}). 
(Notice that in the case of $\mu\rar 3e$ decay the amplitude
is proportional to $g_2^2$ and that's why the decay
probability should be much larger.) 

The amplitude of the proton decay corresponding to Fig. \ref{f-proton} is 
equal to:
\be
A(p\rar l^+ \bar q q) = \frac{\alpha}{\pi}\,
\ln \frac{M_*^2}{m_q^2}\,\frac{ \Lambda^3 R_S^5}{(2\pi)^4}\, 
\psi_l \psi_p \bar\psi_q \psi_q \, .
\label{A-p-lqq}
\ee
It leads to the decay rate: 
\be 
\Gamma(p\rar l^+ \bar q q) = 
\frac{m_p\,\alpha^2}{ 2^{12} \, \pi^{13}}
\left(\ln \frac{M_*^2}{m_q^2}\right)^2 \,
\left(\frac{\Lambda}{M_*}\right)^6 \,
\left(\frac{m_p}{M_*}\right)^{4+\frac{10}{n+1}}\, 
\int_0^{1/2} dx x^2 (1-2x)^{1+\frac{5}{n+1}} .
\label{gamma-p}
\ee
Correspondingly the lifetime of the proton with respect to the inclusive 
decay $p\rar \bar q q l^+ $ is:
\be
\tau_p \approx 10^{29}\,{\rm years}\,
\left(\frac{M_*}{{\rm  TeV}}\right)^{10+\frac{10}{n+1}}
\left(\frac{{\rm  TeV}}{m_p}\right)^{\frac{10}{n+1}-\frac{10}{3}}\,
\left(\frac{100{\rm MeV}}{\Lambda}\right)^{6}\,
\ln^{-2}\left(M_*/\rm TeV\right)\, f^{-1}_p(n),
\label{tau-p}
\ee
where $f_p(n) $ is 1, 1.3 and 2.2 for $n =2,3$ and 7 respectively.

The best experimental lower
bounds at the level $\tau_p >10^{33}$ years~\cite{PDG} 
are established for the modes $p\rar e^+\pi^0$ and $p\rar \nu K^+$. 
For all other 2-body and some three-body modes the bounds are
at the level of  $10^{32}$ years. The disagreement of our result
with experiment can be easily avoided if we take a slightly larger
$M_*$, still even smaller than 3 TeV. On the other hand, we should keep 
in mind that our estimates are by no means rigorous; they
are only true up to an order of magnitude (and possibly some omitted
factors would allow $M_*$ to be as low as 1 TeV). 
For example, if we take into account that the proton inverse size, 
$\Lambda \sim 100 $ MeV is three times smaller than the quark mass
$m_q \sim 300$ MeV, the life-time might become larger by the factor
$(m_q/\Lambda)^6\sim 10^3$.

There is one more argument indicating
that the decay life-time (\ref{tau-p}) may be
underestimated. By assumption, the intermediate virtual BH is a
scalar and since we believe that gravity does not break parity, such 
BH cannot go into a pseudoscalar particle. In other words, the pair $\bar q q$
cannot form a pseudoscalar meson or vector meson.
Thus we come to the important 
conclusion that proton must have predominantly 3-body decay modes: lepton plus 
two mesons with the two mesons in scalar state.
This immediately shifts our estimates for the probability of decay
into ($l^+ +$ 2 mesons) down by an order
of magnitude because of smaller phase space of scalar state.
On the other hand the decay into three leptons is not influenced by this
argument and its life time should be given by Eq. (\ref{tau-p}).
Bearing in mind that the experimental lower bounds on the
proton life-time with respect to
three body lepton channel, $l^+l^+l^-$ ($l=e,\mu$)
are at the level $(8-5)\cdot 10^{32}$ years
we see that proton decays are on the verge 
of experimental discovery if $M_*$ is 
slightly larger than or about 2 TeV. 

There are quite peculiar signatures specific to the model of proton 
decay considered here.  First, as we have already mentioned, the decays
should be mostly 3-body ones. Second, the final state particles must
always contain a positron, $e^+$, or a positive muon, $\mu^+$. 
The branching ratio into three lepton channel is predicted to be larger 
than that into $e^+$ ($\mu^+$)
and two mesons, because it is natural to expect
that the probability of BH decay into a neutral combination of
two leptons (or antileptons, or lepton and antilepton) is more or 
less the same as the probability of the decay into two quarks,
while the probability of the subsequent quark transformation into 
two mesons is smaller than one, because other channels are open.
The energy spectrum of emitted leptons (with the same charge as proton)
is cut-off at higher energies due to the factor $(1-2E/m_p)^{1+10/(n+1)} $,
see Eq. (\ref{gamma-p}).

\begin{center}
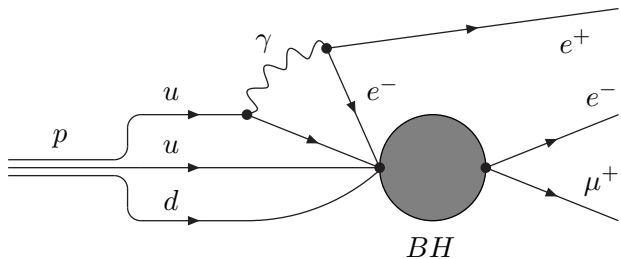
\begin{figure}
\begin{picture}(340,110)(0,-20)
\Line(95,18)(135,18)
\CArc(135,23)(5,270,360)
\Line(140,23)(140,30)
\CArc(145,30)(5,90,180)
\Line(95,15)(145,15)
\Line(95,12)(135,12)
\CArc(135,7)(5,0,90)
\Line(140,7)(140,0)
\CArc(145,0)(5,180,270)
\ArrowLine(145,35)(185,35)
\ArrowLine(185,35)(235,15)
\ArrowLine(145,15)(185,15)
\Line(185,15)(235,15)
\ArrowLine(145,-5)(185,-5)
\CArc(185,70)(75,270,315)
\PhotonArc(215,30)(30,90,170){3}{4}
\ArrowLine(215,60)(325,75)
\ArrowLine(215,60)(235,15)
\ArrowLine(275,15)(325,35)
\ArrowLine(275,15)(325,-5)
\GCirc(255,15){20}{0.5}
\Text(115,26)[]{$p$}
\Text(157,43)[]{$u$}
\Text(237,45)[]{$e^-$}
\Text(157,23)[]{$u$}
\Text(157,3)[]{$d$}
\Text(192,62)[]{$\gamma$}
\Vertex(235,15){2}
\Vertex(275,15){2}
\Vertex(215,60){2}
\Vertex(185,35){2}
\Text(255,-15)[]{$BH$}
\Text(310,63)[]{$e^+$}
\Text(320,43)[]{$e^-$}
\Text(320,10)[]{$\mu^+$}
\end{picture}
\caption{Gravitationally induced proton decay.
Since a 4-body collision is required in order
to form a BH devoid of any quantum number,
the process is strongly suppressed and experimental
constraints can be compatible even with a gravity
scale in the TeV range.}
\label{f-proton}
\end{figure}
\end{center}

\subsection{Neutron-antineutron oscillations \label{ss-n-anti-n}}

Another process where non-conservation of baryons is actively
studied by experiments
is neutron-antineutron transformation. While in many cases, 
as e.g. $SU(5)$ GUT or electroweak theory, 
neutron-antineutron oscillations are impossible
or completely negligible, because they demand change of baryonic 
number by 2, $\Delta B =2$, the
gravitational breaking of global symmetries does
not respect any selection rule and the oscillation time may be
reasonably small. In the framework of the approach presented here, the
neutron-antineutron oscillations are described by the diagram 
of Figure~\ref{f-neutron}.
An estimate of this diagram is very uncertain and the result should be
taken with great caution. There are two loops containing weak
$W$ or $Z$ bosons. Both these loops are logarithmic and if the
ultraviolet cutoff is given by the effective Planck scale, $M_*$,
their contribution is not suppressed as an inverse power of the weak
boson mass. 
Logarithmically divergent part of such loop diagram gives the factor:
\be
\frac{\alpha}{\pi}\ln \left(\frac{\Lambda_{UV}^2}{m_Z^2}\right)
\label{ff}
\ee
where $\Lambda_{UV}$ is the ultraviolet cutoff, which is probably 
reasonable to take equal to the effective Planck mass, $\Lambda_{UV}=M_*$. 
The amplitude of neutron-antineutron transformation contains this factor 
squared.
 
The other part of the diagram, containing the lepton loop, 
is linearly divergent after we perform
integration in the loops containing weak bosons. The
integral should be cut-off at the same energy scale as the neighboring
triangle diagrams with weak bosons, i.e. at $M_*$. In other words, the 
linear divergent part as usually vanishes and the integral is proportional
to the external momentum which, in this case, is of the order of $M_*$.
(See discussion in Sec. \ref{s-lepton}, two paragraphs below 
Eq.~(\ref{sigma-e-mu})). 

As a result of this rather frivolous estimate we obtain for the amplitude 
of the transition of three quarks into three antiquarks:
\be
L_{\Delta B=2} = \left[\frac{\alpha}{\pi}\ln 
\left(\frac{M_*^2}{m_Z^2}\right)\right]^2\, M_*\, \left(R_S^4 E\right)^2
\left( \psi C \psi\right)^3 , 
\label{l-deltaB2}
\ee
where the first factor comes from two triangle parts of the diagram,
Eq. (\ref{ff}), the second factor $M_*$ is the ultraviolet cut-off
of the linearly divergent loop with three lepton lines and one 
virtual BH and the next factor is the coupling constant of BH
to 4 particles. The last term is the product of 6 quark operators 
and $C$ is the charge conjugation matrix.

Now it is straightforward to obtain the time of neutron-antineutron 
oscillations taking the matrix element 
$\langle n| L_{\Delta B=2} |\bar n\rangle$. Since the 
effective energy cutoff in 3$q$-transition into neutron  
is the QCD scale, $\Lambda$, we obtain:
\be
\tau_{n\bar n}= \left[\frac{2\alpha}{\pi}\ln 
\left(\frac{M_*}{m_Z}\right)\right]^{-2} 
\left(\frac{M_*}{\Lambda}\right)^{7+\frac{8}{n+1}}
\,\Lambda^{-1} 
\label{tau-nn}
\ee
(note that we have omitted here the huge loop counting factor $(2\pi)^8$,
because the result is weak anyhow).
With $n=2$ and $M_* \sim 1$ TeV and $\Lambda = 100$ MeV
it corresponds to an oscillation time of
about $3\cdot 10^{19}$ s which is twelve-thirteen
orders of magnitude below the existing experimental
limit: direct searches for $n\rightarrow\bar{n}$ processes
using reactor neutrons put the upper limit on the mean time of
transition in vacuum~\cite{PDG}:
\be
\tau_{n\bar{n}} > 8.6 \cdot 10^7 \: \textrm{s}, 
\label{tau-n}
\ee
while the limit found from nuclei stability is slightly stronger:
\be
\tau_{n\bar{n}} > 1.3 \cdot 10^8 \: \textrm{s} .
\label{tau-n-nucl}
\ee
If the theoretical prediction of Eq. (\ref{tau-nn}) were true, 
the chances to observe 
$(n-\bar n)$-oscillations in the reasonable future are negligible.
However, one can obtain much more optimistic
predictions if there exist supersymmetric partners of the
usual particles, as considered in the following subsection.

\begin{center}
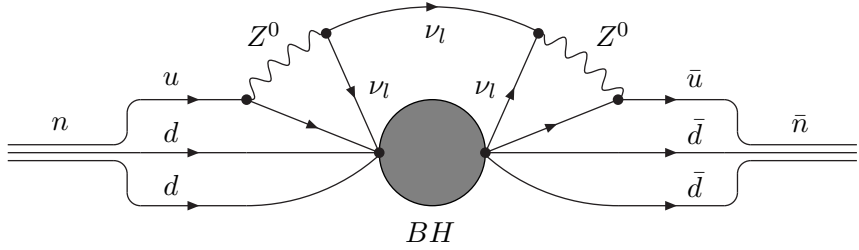
\begin{figure}
\begin{picture}(340,110)(0,-20)
\Line(50,18)(90,18)
\CArc(90,23)(5,270,360)
\Line(95,23)(95,30)
\CArc(100,30)(5,90,180)
\Line(50,15)(100,15)
\Line(50,12)(90,12)
\CArc(90,7)(5,0,90)
\Line(95,7)(95,0)
\CArc(100,0)(5,180,270)
\ArrowLine(100,35)(140,35)
\ArrowLine(140,35)(190,15)
\ArrowLine(100,15)(140,15)
\Line(140,15)(190,15)
\ArrowLine(100,-5)(140,-5)
\CArc(140,70)(75,270,315)
\PhotonArc(210,-20)(90,118,142){3}{4}
\PhotonArc(210,-20)(90,38,62){3}{4}
\ArrowArcn(210,-20)(90,118,62)
\ArrowLine(170,60)(190,15)
\ArrowLine(230,15)(250,60)
\ArrowLine(230,15)(280,35)
\ArrowLine(280,35)(320,35)
\Line(230,15)(280,15)
\ArrowLine(280,15)(320,15)
\CArc(280,70)(75,225,270)
\ArrowLine(280,-5)(320,-5)
\GCirc(210,15){20}{0.5}
\CArc(320,30)(5,0,90)
\Line(325,30)(325,23)
\CArc(330,23)(5,180,270)
\Line(330,18)(370,18)
\Line(320,15)(370,15)
\CArc(320,0)(5,270,360)
\Line(325,0)(325,7)
\CArc(330,7)(5,90,180)
\Line(330,12)(370,12)
\Text(70,26)[]{$n$}
\Text(112,43)[]{$u$}
\Text(191,39)[]{$\nu_l$}
\Text(112,23)[]{$d$}
\Text(112,3)[]{$d$}
\Text(147,60)[]{$Z^0$}
\Vertex(190,15){2}
\Vertex(230,15){2}
\Vertex(170,60){2}
\Vertex(250,60){2}
\Vertex(140,35){2}
\Vertex(280,35){2}
\Text(210,-15)[]{$BH$}
\Text(212,60)[]{$\nu_l$}
\Text(231,39)[]{$\nu_l$}
\Text(310,43)[]{${\bar u}$}
\Text(310,23)[]{${\bar d}$}
\Text(310,3)[]{${\bar d}$}
\Text(279,60)[]{$Z^0$}
\Text(350,26)[]{${\bar n}$}
\end{picture}
\caption{($n - \bar n$)-oscillation mediated by a virtual BH.
If we consider only Standard Model particles, the effect
is negligible.}
\label{f-neutron}
\end{figure}
\end{center}

\subsection{Supersymmetric extension \label{ss-susy}}

Some  of our  estimates, such as for  ($n - \bar n$)-oscillations, 
would be different if there exist supersymmetric partners of the
standard model particles. Since supersymmetry remains a 
hypothesis, not yet proven by experiment, and nothing is known about
the masses of superpartners, except for lower limits on their values,
more ambiguity is introduced into the calculations. Consequently we
here consider anomalous processes with inclusion of superpartners 
as a separate subsection.

The spins of the SUSY particles may differ by 1/2 from their
standard model partners, with
all other quantum numbers being the same. In particular,
there could be scalar quarks (s-quarks) or spin-1/2 partners of
vector bosons mediating interactions. Existence of new types of
elementary particles would modify both neutron-antineutron
transformation and proton decay.

In this case, one of the quarks in the neutron can emit
a neutralino, $\chi^0$, and become a squark, $\tilde q$. This $\tilde q$,
together with remaining quarks, can form a neutral and 
spinless BH. This BH in turn may decay into two antiquarks, $2\bar q$, 
and anti-squark, $\bar{\tilde q}$. The latter captures $\chi^0$ and becomes 
the usual antiquark,
$\bar q$. This completes the transformation of three quarks into 
three antiquarks (see Fig. \ref{f-susy}). 
The analogous process with emission of 
a gluino does not help, because after this emission the remaining two
quarks and one squark become colored.

To find the amplitude corresponding to this diagram we need to calculate
the contribution from the loop containing two s-quark propagators, one 
neutralino propagator, and, most problematic, a propagator of the virtual 
BH, about which we do not have much information. Possibly the result
would be less ambiguous if we were to 
use the old non-covariant perturbation theory
with particles on mass shell with positive energies. The last condition
is important for definition of the vertex of interaction of BH with
particles entering into it or emitted by it. Because of that, the total
energy of incoming or outgoing particles is always of the order of the 
energy of the initial state and the mass of BH is of the same order.
The necessity to use non-covariant perturbation theory 
in our description of interaction of particles with
virtual BHs leads to breaking of Lorentz invariance. 
Another argument in favor of non-covariant perturbation theory
is the impossibility to make Wick rotation with virtual BH -- at least 
we do not know how to do that.

The effective Lagrangian corresponding to the diagram in 
Fig. \ref{f-susy} is the following:
\be
L^{susy}_{\Delta B=2} = \frac{\alpha\, g_3^2}{2\pi}\, 
\frac{m_{\chi}}{m_{SUSY}^2 M_{BH}^2} 
\left( \psi C \psi\right)^3 , 
\label{l-deltaB2-susy}
\ee
where $g_3 = R_S^2 M_{BH}$ is the coupling constant of BH to 3 particles,
two of which have spin 1/2 and one is scalar, 
$\alpha \sim 0.01$ is the electroweak coupling 
constant at characteristic SUSY scale;
$m_{\chi}$ is the mass of neutralino,
$m_{SUSY}$ is the mass of other superpartners, and we assume that they are all 
of the same order of magnitude, $m_{\chi}\sim m_{SUSY}$.
A subtle point is the value of the BH mass. According to the arguments
presented above we take it to be of the same order of magnitude as the  
energies of the external particles;
again, in the absence of any fundamental theory describing 
behavior of virtual BH, we can consider this at best 
a plausible assumption.

Taking the matrix element of operator (\ref{l-deltaB2-susy}) 
between $n$ and $\bar n$ states we find:
\be
\langle \bar n| L^{susy}_{\Delta B=2} |n \rangle =
\frac{\alpha}{2\pi}\, 
\frac{\Lambda^6}{\left(2\pi\right)^8\,m_{SUSY}\,M_*^4}\,
\left(\frac{M_{BH}}{M_*}\right)^{\frac{4}{n+1}} .
\label{nn-susy}
\ee
Here $\Lambda$ is the effective energy which enters in calculating of 
the matrix element 3 quark operators over the neutron state, it is usually
taken about 100 MeV; and $(2\pi)^8 $ is the loop counting factor -- there are
4 loops with virtual quarks which go either into $n$ or $\bar n$ and each
loop provides with $1/(2\pi)^2$.

Taking all the factors together we find for the time of 
$(n-\bar n)$-oscillations: 
\be
\tau_{n\bar n} \approx 3\cdot 10^9\,
{\rm sec}\,
\cdot 10^{\frac{12}{n+1} -4}\,
\left(\frac{100 \,{\rm MeV }}{\Lambda}\right)^6\,
\left(\frac{m_{SUSY}}{300{\rm GeV }}\right)\,
\left(\frac{{\rm GeV }}{M_{BH}}\right)^{\frac{4}{n+1}}\,
\left(\frac{M_{*}}{{\rm TeV }}\right)^{\frac{4(n+2)}{n+1}} .
\label{tau-nn-susy}
\ee
This result looks quite promising. If  $M_*$ is not too much larger than 1 TeV
and the SUSY partners are not far from 300 GeV, the chances to observe 
neutron-antineutron transformations are very good.
According to the model presented here their observation would indicate not
low scale gravity but also low energy SUSY; but this is probably
too far fetched.

We also note that the contribution of SUSY partners to proton decay
is negligible.

\begin{center}
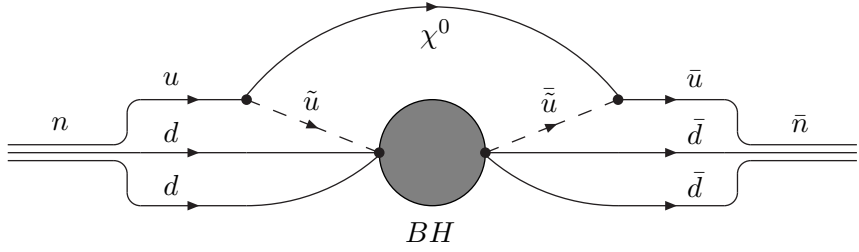
\begin{figure}
\begin{picture}(340,110)(0,-20)
\Line(50,18)(90,18)
\CArc(90,23)(5,270,360)
\Line(95,23)(95,30)
\CArc(100,30)(5,90,180)
\Line(50,15)(100,15)
\Line(50,12)(90,12)
\CArc(90,7)(5,0,90)
\Line(95,7)(95,0)
\CArc(100,0)(5,180,270)
\ArrowLine(100,35)(140,35)
\DashArrowLine(140,35)(190,15){5}
\ArrowLine(100,15)(140,15)
\Line(140,15)(190,15)
\ArrowLine(100,-5)(140,-5)
\CArc(140,70)(75,270,315)
\ArrowArcn(210,-20)(90,142,38)
\DashArrowLine(230,15)(280,35){5}
\ArrowLine(280,35)(320,35)
\Line(230,15)(280,15)
\ArrowLine(280,15)(320,15)
\CArc(280,70)(75,225,270)
\ArrowLine(280,-5)(320,-5)
\GCirc(210,15){20}{0.5}
\CArc(320,30)(5,0,90)
\Line(325,30)(325,23)
\CArc(330,23)(5,180,270)
\Line(330,18)(370,18)
\Line(320,15)(370,15)
\CArc(320,0)(5,270,360)
\Line(325,0)(325,7)
\CArc(330,7)(5,90,180)
\Line(330,12)(370,12)
\Text(70,26)[]{$n$}
\Text(112,43)[]{$u$}
\Text(112,23)[]{$d$}
\Text(112,3)[]{$d$}
\Text(165,35)[]{$\tilde{u}$}
\Vertex(190,15){2}
\Vertex(230,15){2}
\Vertex(140,35){2}
\Vertex(280,35){2}
\Text(210,-15)[]{$BH$}
\Text(212,60)[]{$\chi^0$}
\Text(310,43)[]{${\bar u}$}
\Text(310,23)[]{${\bar d}$}
\Text(310,3)[]{${\bar d}$}
\Text(255,35)[]{$\bar{\tilde{u}}$}
\Text(350,26)[]{${\bar n}$}
\end{picture}
\caption{($n - \bar n$)-oscillation with supersymmetric
particles. In this case the observation of the phenomenon 
may be accessible to future experiments.}
\label{f-susy}
\end{figure}
\end{center}

\section{Heavy quark decays \label{s-heavy}}

We discuss here decays of heavier quarks or mesons containing such
quarks. Though the experimental accuracy in their decays are much
lower than for proton or muon, the effects may be amplified
because of larger masses and it is quite probable that the manifestation
of low scale gravity will be first observed in such decays.

Let us start from the heaviest, the
$t$-quark. Since it has a very large mass, $m_t = 175-180$ GeV, 
close to the assumed gravity scale,
one can expect that B-nonconserving decays of the $t$-quark may be
noticeably enhanced. A B-nonconserving decay $t\rar 4q +l$ is
described by a diagram of the type presented
in Fig.~\ref{f-top}.  The same considerations as those presented in
Sec. 5 lead to the decay rate:
\begin{eqnarray}
\Gamma \sim \epsilon_5
\,\left(\frac{\alpha_{QCD}}{2\pi}\right)^2
\Big(\frac{m_t}{M_{\ast}}\Big)^{10+\frac{10}{n+1}} m_t,
\label{gamma-b}
\end{eqnarray}
where $\epsilon_5 \sim 10^{-10} $ is the 5-body phase space
normalized to the $t$-quark mass.

The total decay rate of the $t$-quark is known to be:
\begin{eqnarray}
\Gamma_{tot} \approx \alpha m_t \approx 7\cdot 10^{-3} m_t .
\label{gamma-tot}
\end{eqnarray}
Using Eqs.~(\ref{gamma-b}) and (\ref{gamma-tot})
we obtain the branching ratio
(always assuming $M_{\ast} \sim 1$ TeV):
\begin{eqnarray}
BR_{\Delta B \neq 0} \sim
10^{-20} - 10^{-19}
\label{BR-B}
\end{eqnarray}
for $n$ between 2 and 7. Probably the decay rate (\ref{gamma-b}) is 
underestimated because of too large phase space suppression factor
and the branching ratio is somewhat larger. In particular it would be larger 
for B-nonconserving decays of mesons or baryons containing $t$-quark
because phase space suppression in this case would be much milder. On the
other hand, the loop counting factor, omitted above, would play its destructive 
role.

At the present time there are no experimental
restrictions on nonconservation of the baryonic number in decays of
particles containing $t$-quark. However, a noticeable baryonic charge
non-conservation in $t$-quark decays and some other anomalously enhanced 
decays of mesons containing $t$-quark may be accessible to future
experiments.

More realistic possibility may be in the leptonic sector:
for example, the BH in Fig. \ref{f-top} can emit an
$e^{\pm}\mu^{\mp}$-pair, leading to the violating the family lepton number
decay $t\rar u e\mu$. Its amplitude is
\be
A(t\rar u e\mu) = \frac{\alpha_{QCD}}{\pi}\,\ln\left(\frac{M_*}{m_t}\right)^2\,
R_S^2 \,\bar\psi_u \psi_t \bar\psi_\mu \psi_e .
\label{a-tuemu}
\ee
The predicted decay rate is
\be
\Gamma(t \rightarrow ue\mu) =
\frac{\alpha^2_{QCD}\, m_t}{16\pi^5}\, \ln^2\left(\frac{M_*}{m_t}\right)^2\,
\Big(\frac{m_t}{M_{\ast}}\Big)^{4+\frac{4}{n+1}}
\int_0^{1/2} dx x^2 (1-2x)^{1+\frac{2}{n+1}}
\ee
and, for $n = 2$, we expect a BR at the level of $10^{-9}$.

On the other hand, $t$-quark may be not the best candidate for search of 
anomalous decays induced by virtual BH because $m_t >m_W$ and the total decay 
width of $t$-containing particles  
is quite large due to the open channel into $W $ and $b$-quark.

Probably a better place to search for low scale gravity effects 
could be decays of $b$-quark or, to be more precise, decays of mesons
containing $b$-quark, in particular $B$-mesons. On one hand, $b$-quark
is lighter than $t$-quark, its mass is about 4.5 GeV, 
and this makes the effects weaker. On the
other hand, the total decay widths of particles containing $b$-quark
are smaller than those with $t$-quark (because $b$ is not heavy
enough to decay into $W$ boson and lighter quarks) and this
would enhance branching ratio of anomalous decays induced by gravity.

As an example, let us consider a decay of $B^0$-meson consisting of a
heavy $b$-quark and light $\bar d$-quark.  
As is known from QCD such a system of light and 
heavy quark has the size of the order of $\Lambda^{-1}$ and the 
characteristic energy of the light quark about $\Lambda$. 
Its decay into two light leptons can be considered in the same way as muon 
or $K$-meson decays
in Sec. \ref{s-lepton}. One should expect that an uncharged 
and nonrotating virtual BH
can be formed directly in collision of $b$ and $\bar d$ and it is 
not necessary to invoke any other virtual particles. 
If this is true (but remember that BH should probably be a scalar and
not a pseudoscalar), then the amplitude of $B$-meson decay into two quarks or two 
leptons is given by the expression
\be
A(B\rar ll) = \frac{g_2^2\,g_{Bqq}}{M_{BH}^2}\frac{4}{(2\pi)^4}
\int \frac{ d^4 p\, (p p_1 + m_b m_q)}{(p^2-m_b^2)(p_1^2-m^2_q)}\, \bar\psi_l \psi_l, 
\label{a-bll}
\ee
where $g_{Bqq} $ is the coupling constant of $B$-meson to two quarks and
 $p_1= p - p_B $.

The cut-off of this integral is determined by the strong interaction scale
$\Lambda$ and could be described by a formfactor:
\be 
F_1= F(p_B^2/m_B^2, p_b^2/m_B^2, p_q^2/m^2_q)
\label{form-fact}
\ee
which vanishes if the participating particles are too far from the mass shell.
However, it is rather complicated to impose this condition in the standard form
of the Feynman integral (\ref{a-bll}). We will use to this end the 
on-mass-shell representation of the Green's functions:
\be
G({\bf x},t) \sim \int \frac{d^3 p}{(2\pi)^3E} \exp \left(-iE t + i{\bf px} \right),
\label{G-mass}
\ee
where $E = \sqrt{p^2+m^2}$. The one loop diagram is now described by the
expression
\be
 \int\frac{d^3 p\, F_2 ({\bf p^2}) }{E_q\,E_b\,(E_B-E_q-E_b)}\,
\left( m_b^2 - p_B p_b + m_b m_q \right) .
\label{one-loop}
\ee
The last term in brackets comes from the fermionic trace and contains the product of 
four-momenta of the virtual fermions. The new formfactor $F_2$ is a function of 
three-momentum ${\bf p^2}$ because all the particles are on-mass-shell but 
energy is not conserved in each vertex. This form-factor is determined by the 
interaction potential and is cut-off at ${\bf |p|} \sim \Lambda$.

This one-loop integral (\ref{one-loop})
can be easily estimated giving the result $\sim \Lambda^2$.
An important fact is that it does not contain the mass of heavy quark, $m_b$ in the
denominator. 

The decay width of $B$-meson into the channel $B\rar ll$ or $B\rar \bar q q$ can now
be estimated as
\be
\Gamma (B\rar ll) \approx 
\frac{m_B\, g^2_{Bqq}}{2^3\,\pi^5}\,
\left(\frac{m_B}{M_*}\right)^{4(1+\frac{1}{n+1})}\, \left(\frac{\Lambda}{m_B}\right)^4 .
\label{gamma-b0}
\ee

For ${g}_{Bqq}=1$, $n=2$, $\Lambda = 100$ MeV and $M_* = 1$ TeV 
we find the life-time $\tau_B \sim 3\cdot 10^{-3} $ s. 
The total life-time
of $B^0$ is $1.5\cdot 10^{-12}$ s. Thus the branching ratio of anomalous
decays with the chosen values of the parameters should be about $5\cdot 10^{-10}$. 
This result is below the existing experimental bounds. 
The branching ratios
of $B^0$ into $e^+e^-$, $\mu^+\mu^-$, $e\mu$ are all bounded by approximately
$10^{-7}$. 
We repeat, however, that estimated branching ratio may be true if pseudoscalar 
BHs are allowed.

If the virtual BH must be scalar, and we consider the decay $B\rar 2l$,
then before collapsing into BH
the system $b\bar d$ should emit a light pseudoscalar (PS) meson, $P$,
which is later to be absorbed by the final state leptons, but the
probability of that is very low (it is a weak interaction process).
Thus the 3-body decays should dominate in the same way as found for
$K$-meson decays in Sec. \ref{s-lepton}. 

The amplitude of the decay $B\rar P + 2l$, where $P$ is a light PS-meson,
is determined by the similar loop integral as above with the only change that 
$m_B^2 - 2E_P m_b $ is substituted instead of $m_B^2$. This leads to further
deviation from the mass-shell pole and the integral is suppressed by 
approximately an order of magnitude in comparison with the previous case,
i.e. $ \sim 0.1 \Lambda^2$.
Correspondingly the decay width is equal to
\be
\Gamma(B\rar Pll) \approx \frac{10^{-2} m_B \left(g_{BPS} m_B\right)^2 }{2^5\pi^7}\,
\left(\frac{\Lambda}{m_B}\right)^4\,
\left(\frac{m_B}{M_*}\right)^{4+\frac{4}{n+1}}\,
\int_0^{1/2} dx\,x(1-2x)^{1+\frac{2}{n+1}} ,
\label{gamma-bpll}
\ee
where $g_{BPS}$ 
is the coupling constant of transition of $B$-meson into PS-meson $P$
and a scalar state of two quarks. It has dimension of inverse mass and it may be 
natural to assume that $g_{BPS}\sim 1/\Lambda$ (this is a quite strong coupling).

For ${g}_{BPS}\sim 1/\Lambda$, $n=2$, $\Lambda = 100$ MeV and $M_* = 1$ TeV the life-time
with respect to this decay is $ 0.2$ s. It leads to the branching ratio 
$BR(B\rar Pll) \approx  10^{-11}$. This relatively large branching is related to a
huge coupling $g_{BPS}\sim 1/\Lambda$.

An interesting process could be a decay of 
$B^0$ into $p+e^-$. The branching 
ratio of this decay should be suppressed with respect to the decay of $B^0$ 
into two leptons (\ref{gamma-b0}) at least 
by the factor $(R_S\Lambda)^6 \lesssim 10^{-24}$, because now
the BH emits four particles and therefore the amplitude of the
process is proportional to $g_2g_4/M_{BH}^2$. 
It makes this decay impossible to observe in foreseeable future.

There can be some other $B$-meson decays as well, with violation of baryon or lepton 
numbers or not, where virtual BH could give noticeable contributions and 
an observation of possible anomalously large branching ratios might
indicate on the effects of virtual BHs.

\begin{center}
\begin{figure}
\begin{picture}(340,110)(0,-20)
\ArrowLine(125,25)(165,25)
\ArrowLine(165,25)(225,25)
\GlueArc(195,25)(30,90,180){3}{5}
\ArrowLine(195,55)(290,75)
\ArrowLine(195,55)(225,25)
\ArrowLine(245,25)(290,55)
\ArrowLine(245,25)(290,35)
\ArrowLine(245,25)(290,15)
\ArrowLine(245,25)(290,-5)
\GCirc(235,25){10}{0.5}
\Text(145,15)[]{$t$}
\Text(195,15)[]{$t$}
\Text(165,55)[]{$g$}
\Text(220,47)[]{$\bar{u}$}
\Vertex(165,25){2}
\Vertex(195,55){2}
\Vertex(225,25){2}
\Vertex(245,25){2}
\Text(235,5)[]{$BH$}
\Text(260,77)[]{$u$}
\Text(300,55)[]{$u$}
\Text(300,35)[]{$d$}
\Text(300,15)[]{$s$}
\Text(300,-5)[]{$\nu_e$}
\end{picture}
\caption{Top decay with BH intermediate state
and violation of baryon and lepton numbers.}
\label{f-top}
\end{figure}
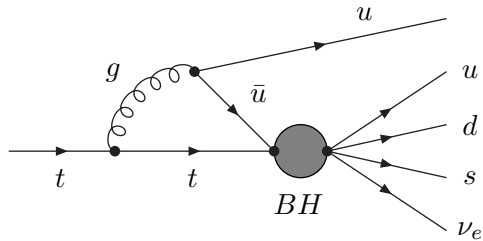
\end{center}


\section{Conclusions \label{s-conclusion}}

Observable effects of gravity in particle physics are an interesting and
fascinating possibility allowed by models with large extra dimensions,
where the fundamental gravity scale can be set in the TeV range. In
such a framework, we have considered rare decays with BH intermediate
states.  Since BH decays violate global symmetries, we expect lepton
and baryon numbers non-conserving processes.  We elevate the classical
requirement of zero charge and zero angular momentum for
sub-Planck-mass black holes to a general conjecture even in quantum
gravity. In that case, the predicted B or L non-conserving decay rates
or the frequency of neutron-antineutron oscillations are interestingly
close to the existing experimental bounds but not in conflict with
them for $M_* > 1$ TeV . It is striking that neutron-antineutron
oscillations and some anomalous decays of $K$ and $B$-mesons are very 
close to the existing experimental bounds and, thus, quite promising 
for observation of TeV gravity effects. 

We have taken $M_*$ as low as possible, namely about 1 TeV.
With higher $M_*$ all experimental bounds can be
easily satisfied, but new physics becomes harder to test.
Unfortunately we do not
know how large can be $M_*$; hence it is impossible to
reject our model by experiment. Even with $M_*$ slightly 
larger than 1 TeV all effects discussed here would be strongly
suppressed.

The model presented here is very speculative and includes plenty of wishful 
thinking. On the other hand, it gives very interesting testable 
predictions with specific signatures absent in other models.
We repeat that with the ``natural'' value of $M_* = 1$ TeV the
predictions for the magnitude of new effects are quite close to 
the existing experimental accuracy.

The way in which virtual BHs are treated here implies some striking 
features, e.g., possible breaking of Lorentz invariance. It may mean, 
in particular, that the magnitude of effects in different coordinate frames
may be significantly different -- a reincarnation of the old concept of the 
ether.


\section*{Acknowledgments}

We thank M. Giannotti , J. Liu, and Yu. Kamyshkov for
useful discussions and suggestions.


\appendix

\section*{Appendix: Short review on current constraints on $M_*$}

To begin with, we stress that, here and throughout
the paper, as fundamental gravity scale we take
the quantity $M_*$, which is related to the
(4+$n$)-dimensional gravitational constant $G_*$ of the
(4+$n$)-dimensional Einstein-Hilbert action by
the simple relation
\be
M_*^{n+2} = \frac{1}{G_*} .
\ee
In the trivial case $n = 0$, we have
$M_*^2 = M_{Pl}^2 = 1/G_N$. In the literature
there are at least other two popular conventions.
Some details on the different possibilities can be
found, for example, in Ref. \cite{giddings}.

In addition to this, we now assume that the
(4+$n$)-spacetime is given by ${\mathcal M}_4 \times T^n$,
where ${\mathcal M}_4$ is the standard 4-dimensional
spacetime we know and $T^n$ is an $n$-dimensional
torus of radius $R$. In this special case, the
volume of the extra dimensions is finite and equal
to $(2\pi R)^n$ and the relation between the
standard 4D Planck mass $M_{Pl} = 1.22 \cdot 10^{19}$ GeV
and the fundamental gravity scale $M_*$ is
\be
M_{Pl}^2 = (2\pi R)^n \; M_*^{n+2} .
\ee
With these two statements in mind, we now review
present experimental lower bounds on the magnitude of
$M_*$. From now on, we follow Ref. \cite{PDG}.

Since a TeV gravity scale allows the emission
of gravitons at colliders \cite{giudice},
constraints on $M_*$ can be obtained looking for
missing energy in processes such as 
$e^+e^- \rightarrow \gamma G$ (the probability that 
the graviton $G$ interacts with the detector is
suppressed by the standard Planck mass $M_{Pl}$,
and therefore negligible). The non-observation
of such events at LEP leads to the 95\% CL 
bounds \cite{lep}
\be
M_* > 1.43, \: 0.76, \: 0.47, \: 0.33, \: 0.25 \:\: {\rm TeV}
\ee
for $n = 2 - 6$.

Much more stringent constraints for $n < 4$ can be
obtained by astrophysical considerations.
We note however that these bounds require the
existence of gravitons lighter than about 100 MeV
and can be evaded if gravitons acquire small extra
contributions to their masses, as suggested 
in \cite{kaloper, dienes, giudice2} (in the scenario
of Ref. \cite{giudice2}, even the $n = 1$ case
cannot be safely excluded). In fact, the astrophysical
environments used to constraint $M_*$ are characterized
by a typical energy per particle of 10 -- 100 MeV
and an effective graviton mass of 100 MeV or more
prevents a copious production of them. These scenarios
leave instead unchanged collider constraints, where
the typical energy is roughly 100 GeV and a 100 MeV
mass is unimportant.

Since gravitons are weakly interactive particles,
a possible their production in a supernova event can compete
with neutrino cooling. Neutrinos detection from SN1987A
requires \cite{hanhart}
\be
M_* \gtrsim 12.5, \: 1.0 \:\:{\rm TeV}
\ee
for $n = 2$ and 3 respectively.

On the other hand, if we consider all gravitons
produced by all the supernovae in the history
of the universe, we can expect a diffuse gamma ray
background due to the graviton decay into photons.
The non-observation of such a diffuse background
by EGRET satellite puts the bound \cite{raffelt1}
\be
M_* \gtrsim 34, \: 2.6 \:\:{\rm TeV}
\ee
for the $n =2$ and the $n = 3$ case.

Finally, noting that gravitons produced in
supernovae events are not high relativistic
particles, we can expect that many of them remain
gravitationally bound to the neutron star
remnant and that their subsequent decay
reheats the surface of the star. The
measured luminosity of some pulsars leads to
the very stringent bounds \cite{raffelt2}
\be
M_* \gtrsim 670, \: 20 \:\:{\rm TeV}
\ee
always for $n = 2$ and 3.



\end{document}